# *A novel non-specular mechanism for chaotic ray scattering of internal waves in 3D anisotropic stadiums*

*Nimrod Bratspiess[1†], Leo R. M. Maas[2], and Eyal Heifet[3]*

[1]*School of Physics and Astronomy, Tel-Aviv University, Tel-Aviv, Israel*
[2]*Institute for Marine and Atmospheric Research Utrecht, Utrecht university, Utrecht, Netherlands*
[3]*Department of Geophysics, Tel-Aviv University, Tel-Aviv, Israel*

Fluids, subject to symmetry breaking by stratification support propagation of anisotropic internal waves (IWs). In the vertical plane, rays representing energy paths obey a non-specular reflection law, as their inclination is solely dictated by their frequency. Although satisfying the linear Poincaré equation, in basins having sloping walls, ray dynamics exhibits nonlinear effects such as convergence onto wave-attractors. In contrast, in the horizontal plane of a basin with vertical walls IWs reflect specularly, and follow chaotic ray paths. Here we present a novel analysis of these competing effects in a 3D IW ray billiard of a stadium having sloping walls. We show and explain how varying the wall's slope, shifts the ray dynamics between regimes of near-ergodicity, chaotic scattering, and non-chaotic scattering with self-similar patterns, despite the basin being closed. The rich results stemming from the interplay between elliptical ergodicity and hyperbolic focusing relate to a broader context of physical phenomena.

*Introduction.*— A point mass, moving frictionlessly in a *closed* stadium-shaped billiard and reflecting specularly from its boundary, generally follows an ergodic path [1–6]. Chaos results from repeated stretching and folding, the result of specular reflections in the channel and at the semi-circles respectively. In *open* systems, specular dynamics may result in chaotic scattering [5,7,8], as demonstrated by the Gaspard-Rice system [9], where the exit angle of a point mass scattered from three circular discs, positioned at the vertices of an equilateral triangle, depends sensitively on the launching angle and impact parameter. In both cases, the path followed by a point mass is identical to that followed by wave rays that satisfy the geometric optics limit of the elliptic Helmholtz equation, $p_{xx} + p_{yy} + (\omega/c)^2 p = 0$, where $p$ is the pressure field, $\omega$ is the frequency and $c$ is the wave velocity. As the Helmholtz equation is invariant under rotations it is implied that wave propagation is isotropic. [10,11]

In contrast, non-specular ray dynamics characterizes internal waves (IWs) in a fluid with a continuous and stably stratified density. Gravity provides both the restoring force for perturbations from equilibrium and renders the fluid anisotropic. In the vertical plane, IWs propagate their energy obliquely with respect to gravity at an inclination dictated by the frequency. This is evident from the dispersion relation (1b) related to the Poincaré equation (1a) [12–14]

$$p_{xx} + p_{yy} - \gamma^2 p_{zz} = 0 \qquad (1a)$$

$$\gamma^2(\omega) \equiv \frac{\omega^2}{N^2 - \omega^2} = \frac{k_x^2 + k_y^2}{k_z^2} = \tan^2\beta \qquad (1b)$$

where $\boldsymbol{k} = (k_x, k_y, k_z)$ is the 3D wave vector, $\beta$ its angle with the vertical, also the angle between energy propagation direction and the horizontal; $N$ is the buoyancy frequency, assumed constant. Hyperbolic dispersion, eq. (1b), relates frequency to wave vector *direction* only, a feature characterizing wave propagation in anisotropic media (e.g. in rotating fluids, plasmas, and metamaterials [15–25]. In these internal waves, the group velocity, signaling energy propagation direction, runs parallel to rays. It is perpendicular to the phase velocity, which is parallel to the wave vector. When reflecting from a wall of slope $s$ at horizontal angle $\phi_{in}$ with respect to the depth gradient, the exit angle $\phi_{out}$ is determined by the reflection law [14,26]

$$\tan \phi_{out} = \frac{(1 - h^2) \sin \phi_{in}}{(1 + h^2) \cos \phi_{in} - 2\,sign(k_{in}^z)h} \qquad (2)$$

where $h \equiv s/\gamma$ denotes the scaled wall slope and $k_{in}^z$ is the incoming ray's vertical wave vector's component. Its sign is positive (negative) for rays propagating downward (upward). If $h > 1$ the bottom slope is super-critical and rays coming from above are focused towards the downslope gradient, $\phi = \pi$. Rays coming from below are defocused reciprocally (Fig. 1b). If $h < 1$ the bottom slope is sub-critical and rays, that can come only from above, are focused towards the depth gradient, $\phi = 0$, (Fig. 1a), or, reciprocally, defocused away from $\phi = \pi$.

Restricting the waves to a single $(x, z)$ vertical plane, the Poincaré equation reduces to a spatial hyperbolic equation in which, by stretching the vertical, $z = \gamma z'$, rays follow characteristics $x \pm z' = const$, resembling the motion of bishops on a chess board. Applied to a 'billiard' with sloping walls these rays generically approach a limit cycle, termed wave-attractor [28,29], characterized by a negative Lyapunov exponent which depends, non-smoothly, on the side wall inclination $\mu$.

---

†bratspiess@mail.tau.ac.il

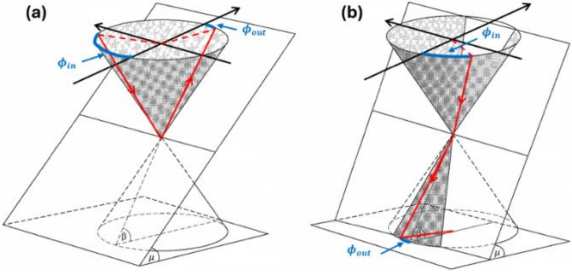

FIG. 1 – (a) Sub-critical internal wave ray reflection. (b) Super-critical internal wave ray reflection. See details in text. Red arrows indicate wave energy propagation direction. Adapted from Fig (2) from [27].

For a 3D stadium – a channel of length $l$, width $w$, and stretched height $\tau$, connected to two semi-circles of radius $w/2$ - having vertical walls, separation of variables reduces the Poincaré equation to the Helmholtz equation, implying specular chaotic ray dynamics in the horizontal plane [1,2,6]. In a 3D stadium with sloping walls of slope $s = \tan \mu$ (all depicted in Fig.2), the Poincaré equation consists of both elliptic horizontal as well as vertical hyperbolic spatial wave operators. As such a basin does not allow separation of variables this raises the question whether chaos or wave attractors will dominate the response [16,29,30], the issue addressed here.

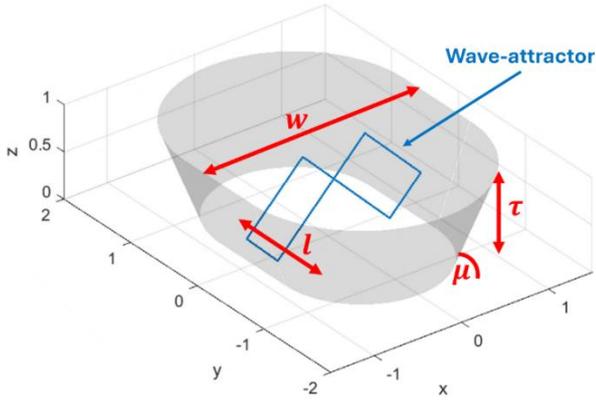

FIG. 2 3D rendering of a linear stadium with $[w, \tau, l, \mu] = [2.2, 1, 0.6, 0.9]$ and a wave-attractor located in a cross section of the channel.

*Numerical experiments.*— For values of $\pi/4 < \mu < \pi/2$ and for a large number $M_c$ of boundary reflections, nearly all ray trajectories approach an attractor that is located at a vertical cross-section of the channel. The $y$-location of such a vertical $(x, z)$ trapping plane, denoted

$-l/2 \leq y_\infty \leq l/2$, depends on the initial launching location and direction. It serves as an exit parameter, despite the stadium being closed. 3D IW propagation inside a stadium can thus be seen as a scattering system. It can be categorized by the trajectory's *convergence time*, $M$, an equivalent measure to the delay time in open scattering systems. If during $M_c$ reflections ray trajectories do not converge onto such a plane, they are classified as *near-ergodic*. Here we neglect the unstable, measure zero part of the invariant set of the mapping, known as whispering gallery modes in the context of 3D IW ray billiards [31–34]

Attractors' structure depends on $\mu$ as demonstrated in Fig.3. We are interested in values of $\mu$ between $\pi/4$ and $\pi/2$. Angles smaller than $\pi/4$ are uninteresting since all sloping walls are sub-critical and rays are attracted to the edges of the basin. Angles between $\pi/2$ and $3\pi/4$ are equivalent to angles between $\pi/4$ and $\pi/2$ up to a mirroring of the basin with respect to the horizontal plane.

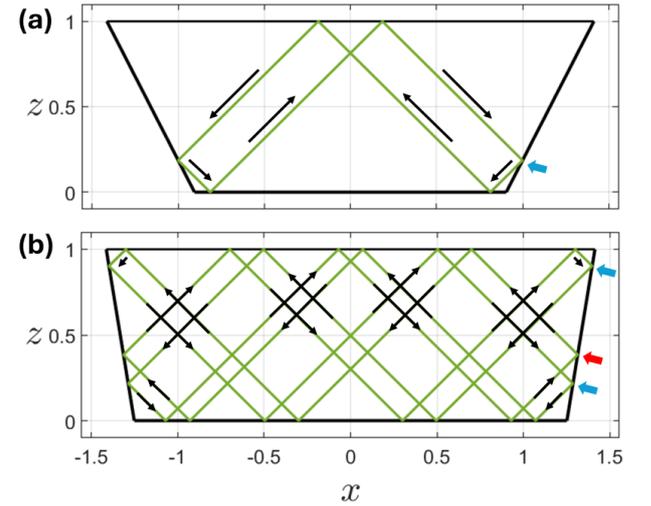

FIG. 3 2D wave-attractors in trapezoid basins with $[w, \tau] = [2\sqrt{2}, 1]$ for (a) $\mu = 1.1$ and (b) 1.41. Black arrows indicate energy propagation direction, blue and red arrows indicate focusing and defocusing reflections respectively.

In the limit $\mu \to \pi/2$ all reflections are specular, and the dynamics is generally ergodic and equivalent to that in 2D specular stadium billiards. In the limit $\mu \to \pi/4$ the sloping walls approach the critical slope where $h = 1$. The reflection law (eq. 1) approaches an absolute focusing for rays coming from above and a practically random scattering for rays coming from below. Consequently, an incident ray coming from above will focus almost perfectly onto a 2D vertical plane. If in subsequent reflections it will come from above, the ray will continue to focus onto a vertical planar wave attractor. However, if in following reflections the ray comes from below, it will scatter and fail to focus onto a planar wave attractor. Therefore, in the limit $\mu \to \pi/4$ rays either converge extremely fast or never, depending on the attractors existing in the vertical cross sections.

Viewing the ray's trajectory as a series of reflections from the boundaries, the convergence time needed to reach the vicinity $\Delta y \ll l/2$ of a trapping plane $y_\infty$, can be evaluated as the number of times, $M < M_c$, a ray hits the boundaries before being trapped. As rays, launched from different points and towards different horizontal directions, have different convergence times, we use a Monte Carlo approach to evaluate the *mean convergence time* $\langle M \rangle$ as a function of the wall angle $\mu$.

*Results.*—Mean convergence times of 3D IW ray Monte Carlo simulations in 3D stadiums with $[w, \tau, l] = [2\sqrt{2}, 1, 0.6]$ and $\mu$ varying from $\pi/4$ to $\pi/2$ are presented in Fig. 4 alongside the corresponding 2D Poincaré plot of symmetric trapezoidal basins with the same width and height. One can observe a correlation between windows in which 2D wave-attractors are simple, i.e. each cycle has only a few surface reflection points, and windows in which 3D mean convergence times are low.

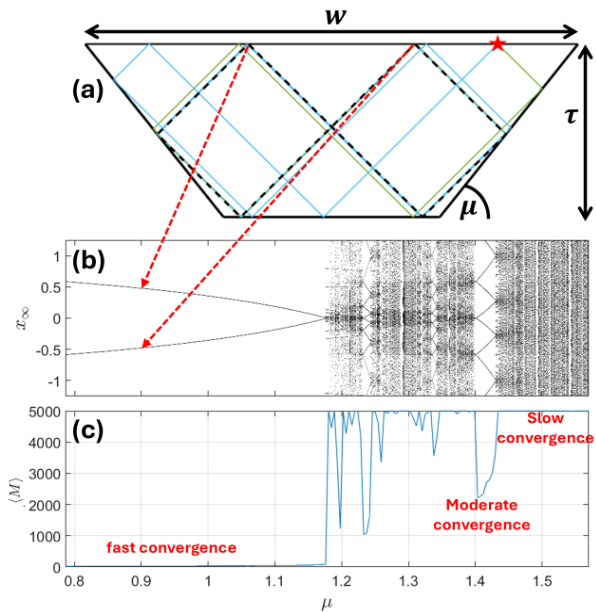

FIG. 4 (a) 2D IWs launched from the red pentagram in a trapezoid basin with $[w, \tau, \mu] = [2\sqrt{2}, 1, 0.9]$. Dashed line denotes attractor. (b) Corresponding Poincaré plot for varying wall angle μ, plotting the last 200 of 1100 surface reflections, a set indicated as $x_\infty$. (c) Mean 3D convergence times, $\langle M \rangle$, for rays reflecting $M_c = 5000$ times in linear stadiums with $[w, \tau, l] = [2\sqrt{2}, 1, 0.6]$ and varying angle μ.

This correlation is explained by the notion of net focusing. The net focusing of an attractor is defined as the number of focusing reflections minus the number of defocusing reflections per total number of boundary reflections (Fig. 3). The simple attractor (Fig. 3a) has two focusing reflections, in which the ray hits the sloping walls from above, out of six total reflections, resulting in a net focusing of 1/3. The complex attractor (Fig. 3b) has four focusing reflections and two defocusing reflections out of twenty-two total reflections, resulting in a net focusing of 1/11. The simple attractor has a higher net focusing than the complex attractor. Therefore, although in both cases the rays focus onto an attractor, focusing onto the simple attractor is much faster, as confirmed in Fig. 4. This is also supported by the Lyapunov exponent being larger in absolute value in the former case of simple attractors [28].

For 3D stadiums with $[w, \tau, l] = [2\sqrt{2}, 1, 0.6]$, windows of moderate convergence appear within regions of near-ergodicity. The widest window of moderate convergence can be seen just below $\mu = 1.43$, equivalent to the slope $s = \tan\mu = 7.05$. As simple attractor windows appear at all values of $\mu$, one should observe moderate convergence windows at higher resolution in $\mu$. Rays launched in basins of any slope converge after some time, for high values of $\mu$ these times diverge, and trajectories appear ergodic as expected. At moderate values of $\mu$, rays converge in times observable in shorter simulations of up to 10,000 reflections. This, however, does not imply that the ray dynamics is simple and stable. Since rays have a significant amount of time to bounce around the basin before converging, they are subject to stretching and folding. Stretching occurs in the channel part of the basin and folding in the semi-circular parts. Stretching is a result of geometric spreading in which order is preserved when channel walls are straight, while folding, resulting from backfolding in the curved semi-circular parts, mixes this order (Fig. 5). One can control the rate of stretching to folding by adjusting the length of the channel. The longer the channel the simpler the ray dynamics. Taking the limit $l \to \infty$ is equivalent of examining a channel with vertical straight end walls replacing the semi-circles. Rays in such a basin experience no folding, and the dynamics is non-chaotic [32,35]. In such a channel, symmetry along the channel direction allows separation of variables in the PDE corresponding to the IW ray dynamics.

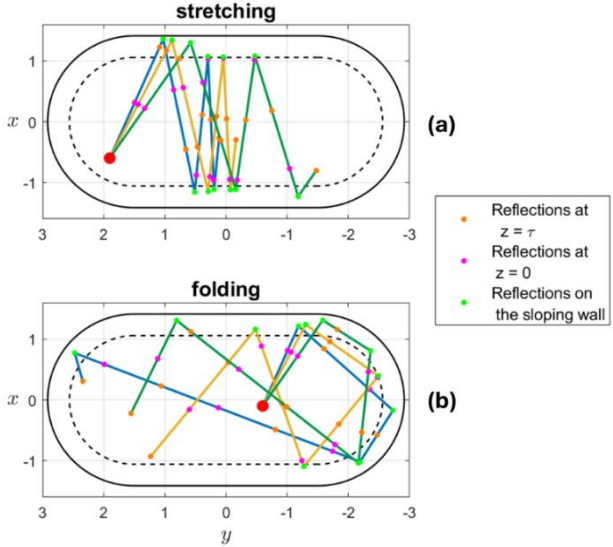

FIG. 5 Top view of stretching and folding of three rays in a linear stadium basin with $[w, \tau, l, \mu] = [2\sqrt{2}, 1, 3, 1.23]$. The rays were launched downwards from the same point on the surface, with initial angles differ by $\pi/50$. (a) In the stretching part of the dynamics rays diverge due to geometric spreading but remain ordered. (b) In the folding part of the dynamics rays get mixed up and order is not preserved.

For moderate values of $\mu$, two rays launched from points extremely close to each other initially de-correlate and end up at unrelated attractors on different vertical trapping planes. By deforming the basin, we change the system from a near-ergodic system for stadiums having near vertical walls (Fig. 6d) to a chaotic scattering system for walls having milder slopes. (Fig. 6a-c). The fact that the set of exit parameters, as the mean convergence time and the convergence plane location, depend sensitively on initial conditions in a closed system suggests a novel mechanism of closed system chaotic scattering. It is interesting to note that IWs are ubiquitous in small scale basins such as lakes and canyons, qualifying as anisotropic due to a stable stratification [36–41].

At values of $\mu$ just above $\pi/4$, 3D basins with cross sections supporting simple attractors suggest fast convergence of rays. Because mid-range convergence times were associated with finite time stretching and folding, resulting in decorrelations of infinitesimally close rays, one expects fast convergence to result in almost no stretching and folding, so that rays remain correlated. To check this conjecture the final trapping plane in cross-channel direction is examined as a function of launching angle $\phi_0$ for rays launched from the same point on the surface (Fig. 7). The trapping plane $y_\infty(\phi_0)$ varies mostly continuously, up to some windows where behavior seems chaotic due to finite resolution in launching angles.

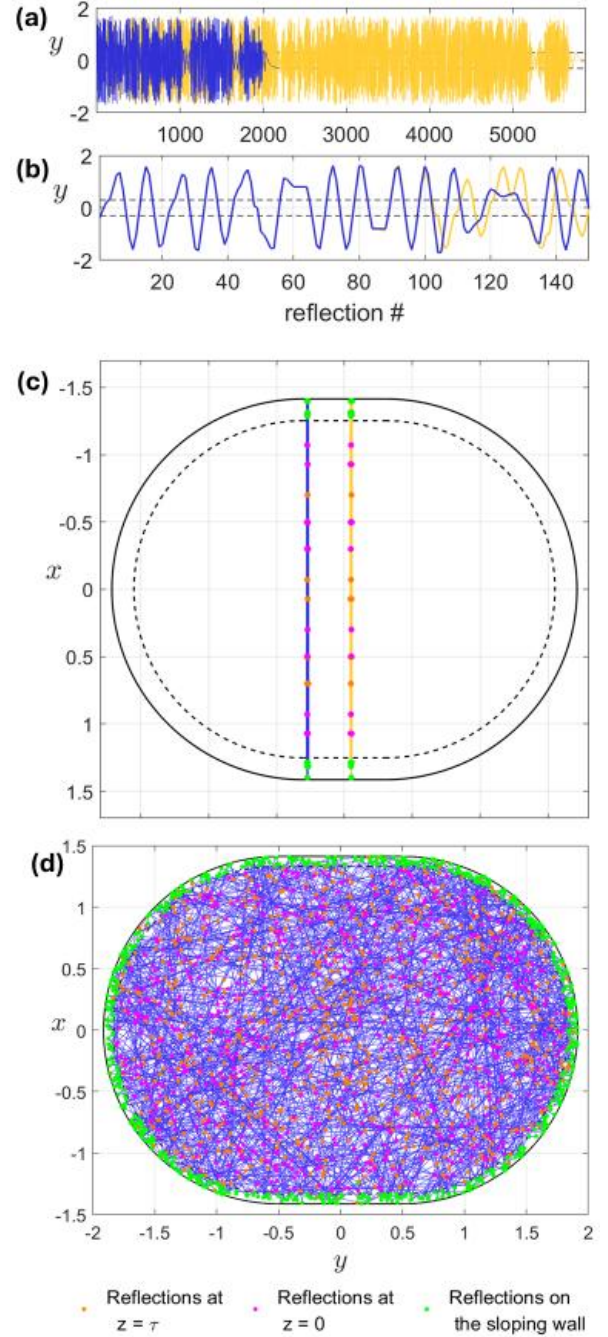

FIG. 6 Convergence plot of two rays launched at a difference of $10^{-8}$ rad in launching angle in a 3D stadium defined by $[w, \tau, l, \mu] = [2\sqrt{2}, 1, 0.6, 0.449\pi]$. (a) y-coordinate of each reflection as a function of reflection number. Each plot ends when the ray is deemed converged. (b) Zoom-in on the first 150 reflections, rays decorrelate after about 100 reflections. (c) Top view of the decorrelated attractors approached by the rays followed in (a), (b). (d) Top view of a near-ergodic trajectory in a 3D stadium defined by $[w, \tau, l, \mu] = [2\sqrt{2}, 1, 0.6, 0.474\pi]$. Each reflection's location is marked by color.

Linear regressions of peak heights against peak widths, $W_i$, in log-space, show that the patterns in Fig. 7 are self-similar for all values of $\mu$. Slopes and correlation coefficients are extremely close to 1. Scaling factors between zoom-ins, $\delta_i = W_i/W_{i+1}$, approach a limit $\delta_\infty$ whose value depends on $\mu$.

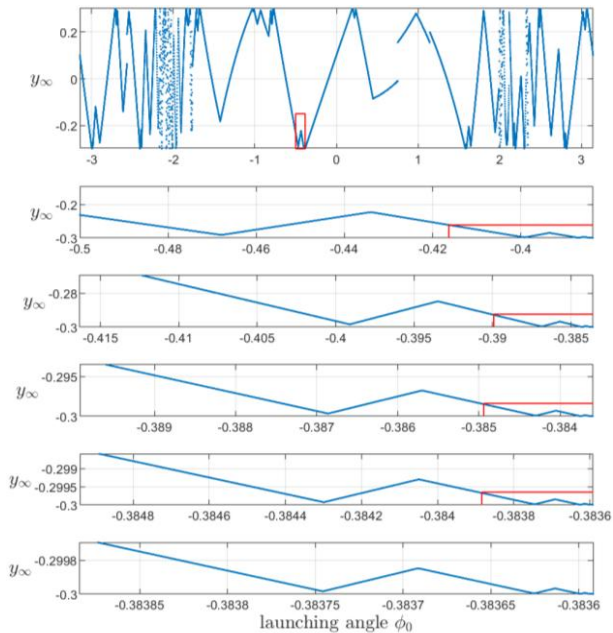

FIG. 7 Final plane of convergence $y_\infty$ as function of launching angle $\phi_0$, defined anticlockwise with respect to the positive x axis. Rays launched downwards from $\vec{r}_0 = (0.7, 0.1, 1)$ in a linear stadium with $[w, \tau, l, \mu] = [2\sqrt{2}, 1, 0.6, 0.318\pi]$. Self-similar patterns in the top panel appear close to the seams, $y_\infty = \pm l/2$, where channel and semi-circles connect. In each panel the pattern in the red box is expanded in the subsequent panel.

*Conclusion.*— Are ray trajectories of internal waves in a 3D anisotropic stadium dominated by chaos or by wave attractors? The surprising answer is both. Nearly all ray trajectories converge onto an attractor, but the location of that attractor and the convergence rates depend sensitively on initial launching position and direction, as well as on basin geometry. The behavior and mean convergence times of internal wave rays in stadiums with different geometries can be predicted by examining their vertical cross-sections, which in turn can be semi-analytically predicted [28]. By decreasing the wall slope, near-ergodicity simplifies to chaotic scattering, and further to a stable scattering that exhibits self-similar patterns depending on mean 3D convergence times. While chaotic scattering is usually found in open systems, such as the Gaspard-Rice system or in heavy ion scattering [5,7–9,42–44] here chaotic scattering is found inside a closed domain. Ironically, this opens a new search for chaotic ray scattering in closed basins, exhibited by different wave agents in different anisotropic media.